\documentclass[%
 prl,%
 twocolumn,
 amsmath,amssymb,
 reprint,%
 preprintnumbers, 
]{revtex4}

\usepackage{graphicx}
\usepackage[colorlinks=true]{hyperref}

\newcommand{\mean}[1]{\mbox{$\langle{#1}\rangle$}}

\begin{document}

\preprint{FERMILAB-PUB-16-559-AD-APC}

\title{Observation of Two-photon Photoemission from Cesium Telluride Photocathodes Excited by a Near-infrared Laser}

\author{H. Panuganti}
\thanks{{Currently at Cornell Laboratory for Accelerator-based Sciences and Education (CLASSE), Cornell University, Ithaca, NY 14853}}
 \email{harsha.panuganti@cornell.edu}
\affiliation{ 
Northern Illinois Center for Accelerator \& Detector Development, and Department of Physics, Northern Illinois University, DeKalb, IL 60115
}%

\author{P. Piot}%
 \affiliation{ 
Northern Illinois Center for Accelerator \& Detector Development, and Department of Physics, Northern Illinois University, DeKalb, IL 60115
}%
\affiliation{%
Accelerator Physics Center, Fermi National Accelerator Laboratory, Batavia, IL 60510
}%

\date{\today}

\begin{abstract}
 We explore nonlinear photoemission in cesium telluride (Cs$_2$Te) photocathodes where an ultrashort ($\sim 100$ fs full width at half max) 800-nm infrared laser is used as the drive-laser in lieu of the typical $\sim 266$-nm ultraviolet laser. An important figure of merit for photocathodes, the quantum efficiency, we define here for nonlinear photoemission processes in order to compare with linear photoemission. The charge against drive-laser (infrared) energy is studied for different laser energy and intensity values and cross-compared with previously performed similar studies on copper [P. Musumeci et al., \emph{Phys. Rev. Lett.}, \textbf{104}, 084801 (2010)], a metallic photocathode. We particularly observe two-photon photoemission in Cs$_2$Te using the infrared laser in contrast to the anticipated three-photon process as observed for  metallic photocathodes. 
 \end{abstract}

\maketitle

Photocathodes are widely used  to generate bright electron bunches with durations comparable to the emission-triggering laser~\cite{jensen}. Semiconductor photocathodes are uniquely attractive due to their high quantum efficiencies (QEs) defined as the number of electrons emitted per unit photon of the drive-laser. Cesium telluride  (Cs$_2$Te) cathodes have substantial QEs (up to 20 $\%$)~\cite{powel,kong}, long lifetimes, and picosecond response times~\cite{jensen}. The photoemission photon energy requirement of Cs$_2$Te dwells around $\sim4$~eV corresponding to an excitation-laser wavelength in the ultraviolet (UV) region. The photoemission band extends from UV to higher wavelengths~\cite{ruth} but any reasonable amount of charge extraction has so far been accomplished using UV laser pulses. Since most of the high-gain lasers use solid state media lasing in the infrared (IR), the typical production of a UV pulse relies on frequency up-conversion from IR to the UV. For laser systems based on the titanium sapphire (Ti:Sapph) medium  (wavelength $\sim800$ nm), the UV pulses needed for photoemission are obtained from frequency tripling of the IR pulses using a two-stage process consisting of a second harmonic generation (SHG) stage followed by a sum frequency generation (SFG) stage. In order to preserve the short pulse duration during the up-conversion process, both stages generally use thin barium borate (BBO) crystals which limits the IR-to-UV conversion efficiency to typically $< 10\%$. 

Operation of UV drive-lasers is technically challenging for high-average-current photoinjectors, while finding cathodes at the longer wavelengths available from solid-state lasers has proven challenging to date~\cite{dowell}. It was pointed out that nonlinear photoemission from metallic photocathodes could be advantageous~\cite{musumeci,li} prompting us to explore possible multi-photon photoemission from Cs$_2$Te using a Ti:Sapph laser.

If an IR laser were to be used for photoemission from Cs$_2$Te, then a nonlinear photoemission process---specifically three-photon photoemission (simultaneous absorption of three photons)---would have to take place if we assume the charge emission is strictly from photoemission and the photoemission threshold is unaltered; this is because, in the current context, the energy of a UV photon equals to that of three IR photons. Multi-photon photoemission has been experimentally explored in metallic photocathodes like copper~\cite{musumeci,muggli,li}, tungsten, molybdenum~\cite{bechtel} but remains unexplored, to our knowledge,  for semiconductor photocathodes like Cs$_2$Te.  This Letter investigates whether Cs$_2$Te exhibits any significant nonlinear photoemission and whether such an effect could have any practical importance with regards to improving the overall photoemission efficiency when compared to the performance of ordinary (single-photon) photoemission.

In a multi-photon process the coefficient of absorption depends on the laser intensity in addition to the photon energy $h\nu$ (where $h$ and $\nu$ are respectively the Planck's constant and photon frequency)~\cite{laud}.

The generalized Fowler-Dubridge (FD) theory for metals describes the emitted current density $J$ as the sum of partial current densities $J_n$ emitted through the corresponding $n$-photon photoemission process as~\cite{fowler,dubridge}     
     \begin{eqnarray} 
     J = \sum_{n}a_n\Big [\frac{e(1-\mathcal{R}_\nu)}{h\nu}I\Big]^nAT^2F(\xi)\equiv \sum_{n}J_n(h\nu) ,\label{eq:FowlerDubridge}
     \end{eqnarray}
where $a_n$ is a material-dependent coefficient, $e$ the electronic charge, $\mathcal{R}_\nu$ the reflectivity of the cathode material, $I$ the laser intensity, $A$ the Richardson's constant, $T$ the mean electronic temperature, and $F(\xi)$ the Fowler function whose argument is  $\xi \equiv \frac{nh\nu-e\Phi}{k_bT}$, where $\Phi$ is the work function, and $k_b$ the Boltzmann constant. The relationhip between $I$ and $J$ in FD theory can still be employed for the case of semiconductor cathodes since the FD theory follows the three-step model~\cite{spicer}. However, the work function is generally replaced by $\Phi=\mathcal{E}_t$ where $\mathcal{E}_t$ is the threshold energy that includes the valence-to-conduction band gap and electron affinity; see, e.g., Ref.~\cite{klaus}.  Considering the partial currents to adopt the form $J_n=\frac{Q_n}{\tau\mathcal{A}}$, where ${\mathcal{A}}$ is the laser-spot area on the photocathode surface and given the laser intensity $I=\frac{\mathcal{E}}{\tau\mathcal{A}}$,  one expects the charge emitted through $n$-photon photoemission mechanism to be $Q_n\propto \frac{\mathcal{\mathcal{E}}^n}{\tau^{n-1}\mathcal{A}^{n-1}}$ where $\mathcal{E}$ is the drive-laser energy, and $\tau$ the laser pulse duration. The latter relationship indicates that $\tau$ and $\mathcal{A}$ play similar roles in the charge emission. Further, we introduce the QE associated to the $n$-order photoemission as 
\begin{eqnarray}\label{eq:QEnphoton}
\eta_n\equiv\frac{h\nu}{e}\frac{dQ_n}{d\mathcal{E}}\propto\frac{\mathcal{E}^{n-1}}{\tau^{n-1}\mathcal{A}^{n-1}}.
\end{eqnarray}           
The right-hand side proportionality implies that for a single-photon photoemission process ($n=1$) the QE is independent of the laser parameters in contrast to the case of a multi-photon photoemission ($n > 1$).
 
Experimentally, one can infer the QE of a photocathode based on the correlation between the charge emitted $Q$ by the given photocathode and the corresponding drive-laser energy $\mathcal{E}$. Usually, the QE is not used to described multi-photon photoemission processes as the ratio $Q/\mathcal{E}$ is not a constant. Here we introduce the IR and UV QEs (in engineering units) as respectively
 \begin{eqnarray} 
\eta_{IR}(\mathcal{E})\mbox{[\%]}&\simeq&0.16 \frac{Q(\mathcal{E}_{IR})[\mbox{nC}]}{\mathcal{E}[\mbox{$\mu$J}]} \mbox{~and} \label{eq:QE_IR}\\
\eta_{UV}\mbox{[\%]}&\simeq&0.47 \frac{Q[\mbox{nC}]}{\mathcal{E}_{UV}[\mbox{$\mu$J}]}= 0.47\frac{Q[\mbox{nC}]}{\kappa \mathcal{E}_{IR}[\mbox{$\mu$J}]}\label{eq:QE_UV},
     \end{eqnarray}
where $\kappa \in [0,1)$ represents the IR-to-UV conversion efficiency, and $\mathcal{E}_{IR}$ and $\mathcal{E}_{UV}$ are respectively the IR and UV laser energies.\\
\begin{figure}[htb]
\includegraphics[height=50mm]{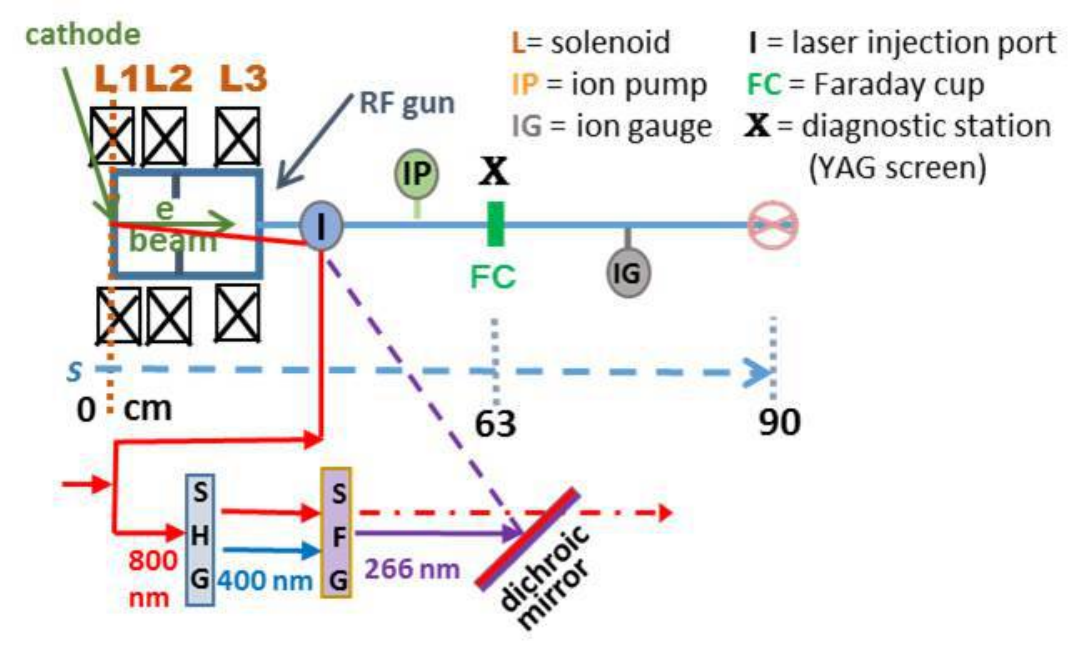}
\caption{\label{fig:schematic} Overview of the HBESL electron source at Fermilab. The photocathode is located at $s=0$~cm and the produced electron beam travels along the $s>0$ direction. The ``SHG" and ``SFG" labels respectively indicate the second harmonic- and sum frequency generation crystals.}
\end{figure}

The experimental investigation of multi-photon photoemission process from Cs$_2$Te cathode reported in this Letter was explored at the high brightness electron source laboratory (HBESL), Fermilab. The HBESL facility consists of a 1.5-cell L-band 1.3-GHz RF
gun ($1+\frac{1}{2}$ cell) powered by a 3-MW klystron. The maximum electric field on the photocathode surface was limited to $E_0 \simeq 26.8$~MV/m in the present experiment. Cs$_2$Te cathodes at HBESL are prepared by first evaporating tellurium onto a hot molybdenum substrate in ultra-high vacuum and then vapor depositing cesium over tellurium while monitoring the QE using an arc lamp. The cesium deposition process is stopped when the QE is maximized~\cite{sertore}. The photocathode drive-laser consists of a broadband (200 nm) {\sc Octavius-85M}
oscillator followed by a Spectra-Physics\textsuperscript{\textregistered} regenerative amplifier. The oscillator is locked to the RF master clock 
with a jitter $<200$~fs. The $\le4$-mJ amplified IR pulse typically has a 100-fs full-width at half max (FWHM) duration. The IR-to-UV SHG and SFG conversion
processes incorporate respectively a 0.300- and 0.150-mm BBO crystals. The crystal thicknesses were optimized with the {\sc snlo} software~\cite{snlo} and short UV pulses (130 fs FWHM) were measured using a polarization gating UV frequency-resolved optical gating technique. The produced pulses were short enough to enable the
production of uniformly charged ellipsoidal bunches using the blow-out regime~\cite{piot}. A schematic of the experimental setup used for our experiment appears in Fig.~\ref{fig:schematic}.

\begin{figure}[b!!!]
\includegraphics[height=0.30\textwidth]{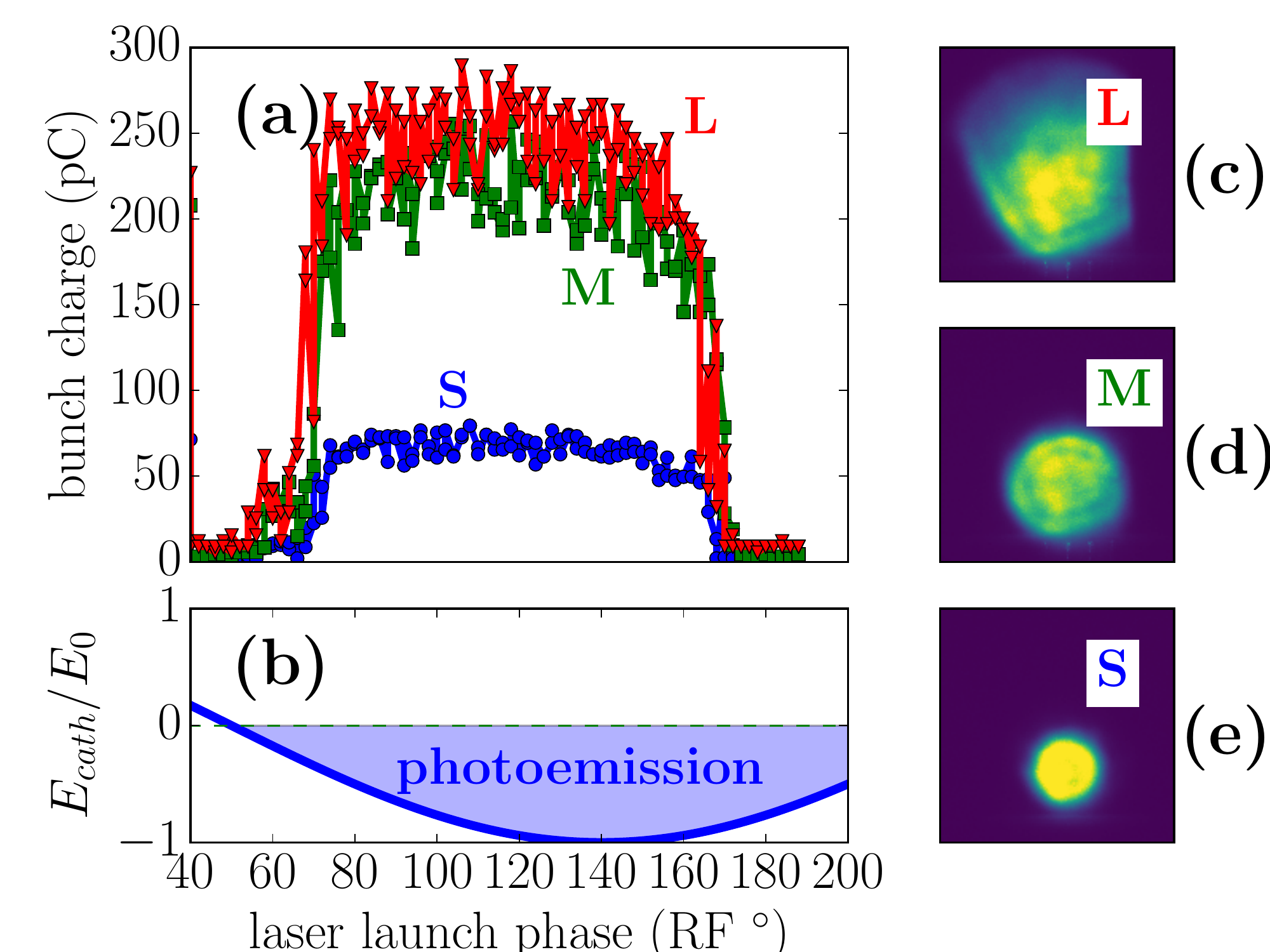}
\caption{\label{fig:chargephase}Charge dependence on the laser--gun relative phase (with an arbitrary phase offset) (a) for three different laser spots on the photocathode with rms sizes 1.72 (trace ``S"), 2.45 (trace ``M") and 3.81~mm (trace ``L"). The corresponding laser transverse-density distributions are displayed in images (c), (d) and (e), each shown over a $1\times 1$~cm$^2$ area. Plot (b) indicates the range of phase where charge extraction can occur, where $E_{cath}$ and $E_0$ are the electric field on the cathode and the peak RF field respectively.}
\end{figure}

In the present experiment the IR pulse was directly sent to the cathode. The laser pulse energy could be measured using a calibrated energy meter located before the optical-injection port; see Fig.~\ref{fig:schematic}. The produced bunch charge downstream of the RF gun was recorded 63 cm downstream of the photocathode using a Faraday cup. During the experiment the ultra-high vacuum pressure was monitored and maintained $< 5 \times 10^{-9}$~Torr as required for extended life time operation of the Cs$_2$Te photocathode. The bunch charge was observed to be strongly dependent on the relative phase between the laser pulse and the gun RF field as summarized in Fig.~\ref{fig:chargephase}(a). The three traces correspond to three different laser transverse distributions on the photocathode shown in Fig.~\ref{fig:chargephase}(c-e). 

\begin{figure}[hhhhh!!!!!!]
\includegraphics[height=0.30\textwidth]{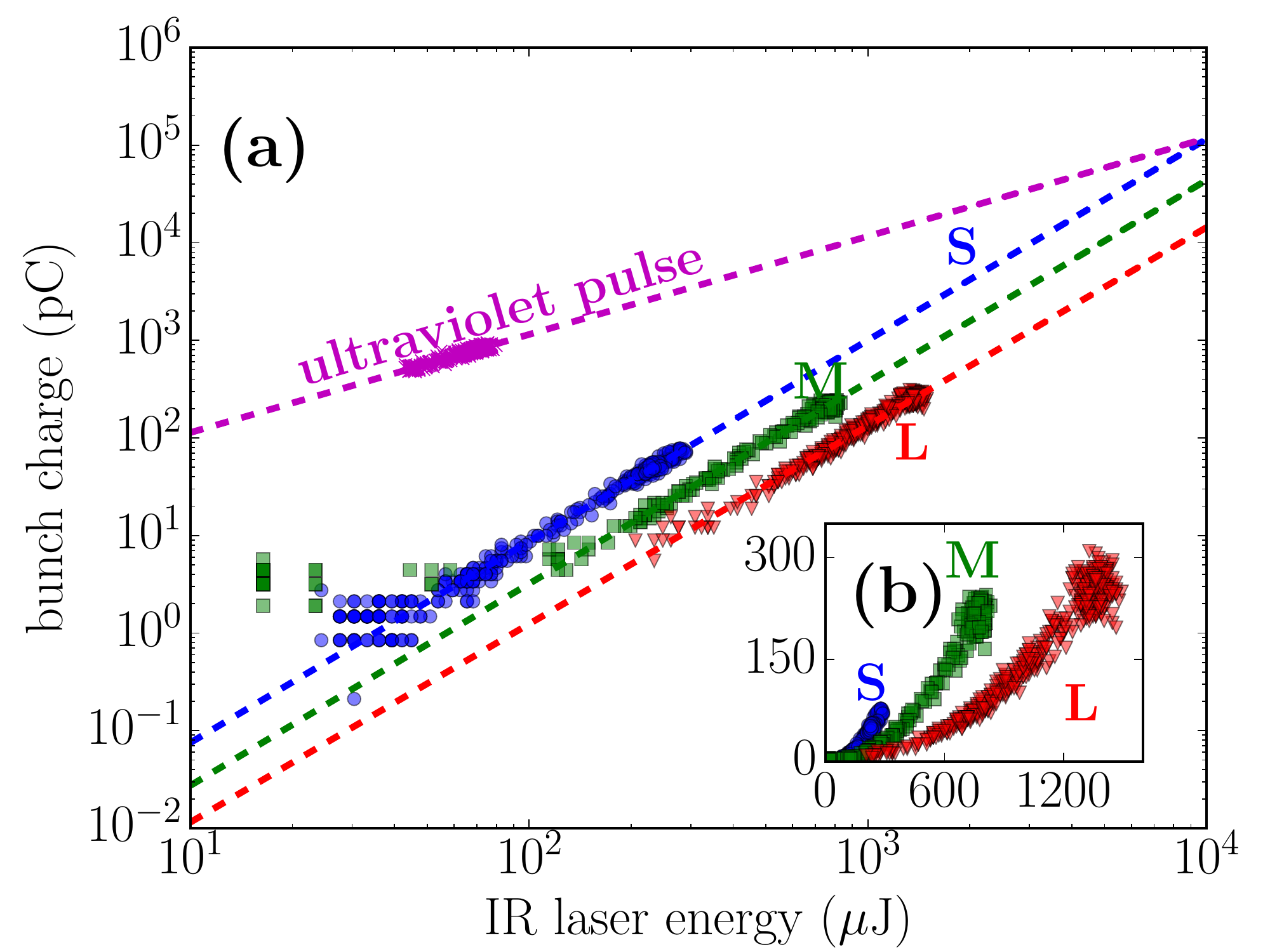}
\caption{\label{fig:IRcharge} Charge dependence on the IR laser energy for the cases of three laser spot sizes (``L", ``M", and ``S") considered in Fig.~\ref{fig:chargephase} (IR pulse on the photocathode) and for the case of an ultraviolet pulse impinging on the photocathode. Inset (b) shows the data corresponding to IR illumination on a linear scale. }
\end{figure}

The maximum charge produced via this IR-pulse triggered emission process was close to 300~pC comparable to what was recently achieved via three- and two-photon photoemission from copper cathodes~\cite{musumeci,li}. To confirm the nature of the multi-photon photoemission process, the charge was recorded as a function of the IR laser energy for the same three different laser spot sizes considered in Fig.~\ref{fig:chargephase} as shown in Figure~\ref{fig:IRcharge}. The nonlinear character of the charge--energy traces reported in Fig.~\ref{fig:IRcharge} [inset (b)] confirms a nonlinear photoemission process. Additionally, a stronger nonlinear dependence of the charge on the laser energy is observed for the smaller laser spot sizes on the cathode. The latter observation confirms that higher laser energy fluxes [J/m${^2}$] produce higher charges via photoemission from the IR laser, as described by Eq.~\ref{eq:QEnphoton}. For comparison we also report the charge measured using UV laser pulse in Fig.~\ref{fig:IRcharge}(a) as a function of the IR pulse energy. For the latter curve, the IR energy is inferred from the IR-to-UV conversion efficiency of $\kappa \simeq 2.5$\% during our experiment. Finally, a regression of the UV data provides the QE to be $\eta_{UV} \simeq 0.22 \pm 0.04$\%.

\begin{table}[hhh!]
\caption{\label{tab:fit} Fitted parameters for charge versus laser energy parametrization $Q \mbox{[pC]}=\beta ({\cal E}_{IR}\mbox{[$\mu$J]})^{\alpha}$ and for the charge density versus optical intensity parametrization $\sigma \mbox{[pC/(mm$^2$)]}=\varrho (I\mbox{[GW/(cm$^2$)]})^{\zeta}$. The ``L", ``M", and ``S" cases corresponds to the ones reported in Figs.~\ref{fig:chargephase} and~\ref{fig:IRcharge}. }
\begin{center}
\begin{tabular}{l c c c c }
\hline \hline  case         &   $\alpha $  & $\log(\beta)$  & $\zeta$ & $\log(\varrho)$ \\
S     &    $2.06 \pm 0.11$      & $-7.33 \pm 0.23$  &    $2.06 \pm 0.11$      & $-4.95 \pm 0.28$\\
M     &    $2.07 \pm 0.11$      & $-8.36 \pm 0.24$  &    $2.07 \pm 0.11$      & $-5.22 \pm 0.25$\\
L     &    $2.03 \pm 0.10$      & $-9.14 \pm 0.23$  &    $2.09 \pm 0.11$      & $-5.37 \pm 0.33$\\
\hline 
UV & $1.01 \pm 0.1$      & $2.41\pm 0.07$  & $1.01 \pm 0.1$     & $2.41\pm 0.07$ \\
\hline \hline
\end{tabular}
\end{center}
\end{table}

To further quantify the dependence of the charge on the IR laser energy, we consider the data represented on a log-log scale in Fig.~\ref{fig:IRcharge}(a). A linear regression of the log-log data indicates a charge scaling as $Q \mbox{[pC]}=\beta ({\cal E}_{IR}\mbox{[$\mu$J]})^{\alpha}$ with $\mean{\alpha}=2.07 \pm 0.19$ (where $\mean{~}$ represents the averaging over the three IR data sets)---see summarized value in Tab.~\ref{tab:fit}---indicating that two-photon photoemission is the dominant process rather than the anticipated three-photon photoemission, based on Eq.~\ref{eq:FowlerDubridge}. Also, the data suggests that the modified photoemission threshold of Cs$_2$Te---or otherwise the emitting material---for such a two-photon photoemission is $<3.1$ eV, the energy corresponding to two photons of 800-nm laser. Some data points at lower charge/energy values with possible higher contribution from background noise were omitted for linear regression. Finally, as expected assuming a constant laser energy, we find that the coefficient $\beta$ increases as the laser spot size decreases.  $\beta$ may not be inversely proportional to the square of the radius of the laser spot envelope for a nonuniform laser intensity distribution like in the current study.

%

\begin{figure}[htb]
\includegraphics[height=55mm]{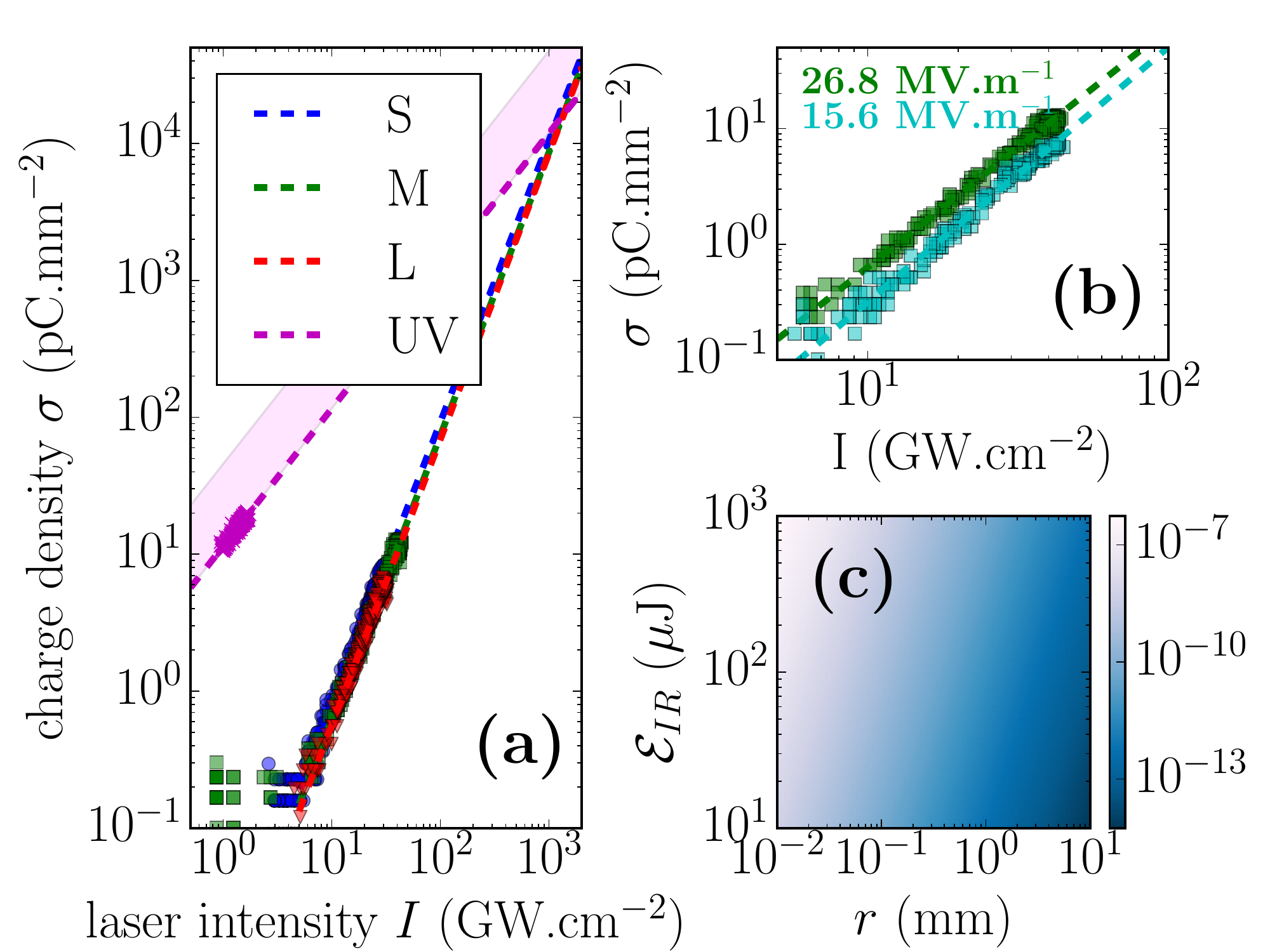}
\caption{\label{fig:JCompare} Charge density evolution as a function of the IR laser intensity (a). The markers represent data (with color coding following convention of Fig.~\ref{fig:IRcharge}) while the dashed lines are the results of linear regression with parameters reported in Tab.~\ref{tab:fit}. The shaded area for the UV data indicates the range of charge density attaintable with $\kappa\in[2.5,10]$~\%. Plot (b) compares the emitted charge density for two cases of operating fields (with values shown as labels) and plot (c) gives the inferred IR QE as function of the laser energy and spot radius ($r$) on cathode.}
\end{figure}

Taking Eq.~\ref{eq:FowlerDubridge} and specializing to the case of a two-photon photoemission process, we find the associated charge density to be $\sigma_2 =\varrho I^{\zeta}$ where $\zeta=2$, $\varrho$ is a material-dependent constant, $I$ the laser optical intensities in units of [W/m$^2$] and the charge density is expressed in unit of [C/m$^2$]. Figures~\ref{fig:JCompare}, compares the emitted charge density as a function of the laser intensity for the three cases of laser spot sizes ``L", ``M", and ``S" on a log-log scale. The data confirms Eq.~\ref{eq:FowlerDubridge} holds for all the three laser spot sizes, i.e all the three linear curves line-up closely and have same intercept with average value $\mean{\log (\varrho)}= -5.18\pm 0.49$ over the three data sets; see also Tab.~\ref{tab:fit}. Over the duration of our experiment the value of $\varrho$ remained constant giving confidence that no change in material properties occurred (e.g. surface contamination or ablation). For comparison the single-photon photoemission process data (obtained by impinging a UV pulse on  the photocathode) is also reported in Fig.~\ref{fig:JCompare}(a) along with its corresponding fit. In comparison with the two-photon photoemission, the UV yield is several orders of magnitude higher than that of the IR emission. The point of intersection between the IR and the UV curves, if extrapolated in Fig.~\ref{fig:JCompare}(a), happens at optical intensity around $I\simeq 1500 \pm 300$~GW/cm$^2$ which indicates that the intrinsic efficiency of IR emission for Cs$_2$Te is well below that of UV emission in the intensity scale of practical importance. This observation is in contrast to the three-photon photoemission from copper studied in Ref.~\cite{musumeci} which suggests that IR emission from copper is preferable to UV emission at intensities  $I> 20$ GW/cm$^2$. This difference stems from the fact that $(i)$ the three-photon process has a steeper slope (three) than two-photon photoemission and $(ii)$ that the UV QE of copper is two orders of magnitude lower than that of the considered Cs$_2$Te cathode; see Ref.~\cite{musumeci}.     

In the presence of large (MV/m) applied electric fields at cathode surface, the effective photoemission potential barrier is lowered resulting in a higher photoemission yield via Schottky effect~\cite{Schottky}. Figure~\ref{fig:JCompare}(b) juxtaposes the evolution of charge density as a function of laser intentisy for two cases of applied electric field $E_0$ considering the case of the medium laser spot size (case ``M"). A decrease in $E_0$ from 26.8 to 15.6~MV/m results in a change in $\log\varrho$ from $-5.22\pm 0.25$ to $-6.21\pm 0.40$ corresponding to a decrease of the charge density by a factor 2.7 thereby confirming the field-dependent character of the photoemission process.  
             
Finally, we use the presented data to infer the IR QE as a function of laser spot size and energies; see Fig.~\ref{fig:JCompare}(c). The QE is inferred as  $\eta_{IR}[\%] \simeq 0.16\times 10^{-3}\alpha\beta{{\cal E}_{IR}[\mu \textrm{J}]}^{\alpha - 1}$ where $\alpha\simeq 2$ and $\beta$ can be readily interpolated for different laser spot sizes from Tab.~\ref{tab:fit}.  As expected, a smaller laser spot size yields a larger $\beta$ and hence a higher value for $\eta_{IR}$ which serves as a figure of merit for the performance of a photocathode for nonlinear photoemission.  \\

%
%
In summary, two-photon photoemission from Cs$_2$Te has been observed with a 800-nm IR laser. It is unclear whether such an emission has any contribution from the bulk Cs$_2$Te material. It is for instance possible that photocathode surface contamination, defects or grain boundaries can play a role in the observed two-photon process. A second Cs$_2$Te photocathode tested also exhibited two-photon photoemission from the IR laser, consistent with the cathode considered in this study. The QE associated with the two-photon process was measured to be several orders of magnitude smaller than that of the linear photoemission from Cs$_2$Te. Such an observation implies  the use of nonlinear photoemission for the generation of high charge using Cs$_2$Te is impractical as the laser energies where the QE from a nonlinear photoemission becomes comparable to that of the ordinary photoemission are above the damage threshold. On the other hand, the data presented could provide an impetus to further explore multi-photon processes in semiconductor cathodes and, e.g., investigate  the definitive band structures of Cs$_2$Te under strong electromagnetic laser fields. Further investigations into the anomalous  IR two-photon photoemission in Cs$_2$Te cathodes could possibly motivate the chemical engineering of new classes of Cs$_2$Te photocathodes that could possibly take advantage of nonlinear photoemission from IR lasers. An example would be to design  a plasmonic Cs$_2$Te photocathode with resonance matched to the IR wavelength~\cite{plamon}. 

This work was supported by the US Department of Energy (DOE) under contract DE-FG02-08ER41532 with Northern Illinois University. FermiLab is operated by the Fermi Research Alliance, LLC under Contract No. DE-AC02-07CH11359 with the US DOE.

\end{document}